\documentstyle[aps,prl,multicol]{revtex}
\begin{document}
\draft
\wideabs{
\title{Measurement of an Anisotropic Energy Gap in Single Plane
Bi$_2$Sr$_{2-x}$La$_x$CuO$_{6+\delta}$}
\author{J. M. Harris,$^1$ P. J. White,$^1$ Z.-X. Shen,$^1$
H. Ikeda,$^2$ R. Yoshizaki,$^2$
H. Eisaki,$^3$ S. Uchida,$^3$
W. D. Si,$^4$ J. W. Xiong,$^4$ Z.-X. Zhao,$^4$
and D. S. Dessau$^5$}
\address{$^1$Department of Applied Physics and Stanford Synchrotron 
Radiation Laboratory, Stanford University, 
Stanford, CA 94305\\
$^2$Institute of Applied Physics and Cryogenics Center, 
University of Tsukuba, Tsukuba, Ibaraki 305, Japan\\
$^3$Department of Superconductivity, The University of Tokyo,
Yayoi 2-11-16, Bunkyo-ku, Tokyo 133, Japan\\
$^4$National Laboratory for Superconductivity, Institute of Physics, Chinese
Academy of Sciences, Beijing 100080, China\\
$^5$Department of Physics, University of Colorado, Boulder, Colorado  
80309-0390 
}
\date{\today}
\maketitle

\begin{abstract}
We report angle-resolved photoemission spectra both above and below $T_c$ in
the single-plane cuprate superconductor Bi$_2$Sr$_{2-x}$La$_x$CuO$_{6+\delta}$. 
The superconducting state measurements show a highly anisotropic
excitation gap with a maximum magnitude smaller than that of 
the bilayer compound Bi$_2$Sr$_2$CaCu$_2$O$_8$ by a factor of 3.  
For a range of doping, the gap
persists well above $T_c$, behavior previously associated
with underdoped bilayer cuprates.  The anisotropy and 
magnitude of the normal-state gap are very similar to the 
superconducting state gap, indicating that the two
gaps may have a common origin in a pairing interaction.
\end{abstract}

}

A central issue in the physics of high-$T_c$ superconductivity
is the role of coupling between the two-dimensional copper-oxygen
planes in producing superconductivity.  The $T_c$ of these materials
tends to increase with the number of layers per unit cell.  It is currently
an open question whether the superconducting state order 
parameter symmetry will be the same in one-layer and 
the more strongly coupled two-layer compounds.  Angle-resolved
photoemission spectroscopy (ARPES) has the potential to resolve
this issue since it is able to
measure directly the anisotropy of the superconducting state
gap (the magnitude of the order parameter).  In the 
two-plane material Bi$_2$Sr$_2$CaCu$_2$O$_{8+\delta}$ (Bi2212), 
the gap was found by ARPES to be highly
anisotropic and consistent with a $d_{x^2-y^2}$ order 
parameter\cite{Shen,Campuzano}.
We report measurements of the one-plane material Bi2201 
that show a similarly large anisotropy with a
smaller overall gap magnitude.

In underdoped Bi2212, ARPES measurements have shown that 
the anisotropic gap persists well above $T_c$\cite{Bi2212ARPES,Harris},
consistent with many other experiments that have shown
a pseudogap or spin-gap in the normal state of cuprate
superconductors\cite{pseudogap}.  Current evidence for 
the normal state gap in
one-plane materials is much weaker than in two-plane
materials, and its existence is controversial\cite{MillisMonien}.  Our 
results show a clear normal state gap up to high temperatures
in optimally-doped and underdoped Bi2201, but not in
overdoped Bi2201.

Single crystal samples of Bi$_{2+x}$Sr$_{2-(x+y)}$La$_y$CuO$_{6+\delta}$ 
were grown using a floating zone method, and for comparison by a
self-flux method.  X-ray scattering confirms that the crystals 
are single-phase Bi2201, and electron-probe microanalysis was used to 
measure the atomic ratios of the cations.  Substitution of trivalent 
La or Bi for divalent Sr reduces the hole
concentration in the CuO$_2$ planes.  The effect of La doping
goes beyond changing the carrier density\cite{Nameki}, however, 
and raises the maximum $T_c$ from 10 K to 30 K.\@  A roughly 
parabolic dependence of $T_c$ on $(x+y)$
has been observed\cite{Yoshizaki}.
Our optimally-doped crystals ($T_c=$29 K) come from 
substituting La=0.35 for Sr.
With no La substitution, Bi/Sr ratios of 2.3/1.7 and
2.1/1.9 give the underdoped samples with $T_c<$ 4 K and the 
overdoped samples with $T_c =$ 8 K, respectively.
The transition temperatures were taken as the 
zero resistance values and confirmed by
SQUID magnetization measurements.  The transition
widths are less than 2 K.  The resistivity curves
give linear-$T$ behavior for the optimally-doped samples\cite{Yoshizaki}, 
positive curvature for the overdoped samples typical of 
other overdoped cuprates, and linear-$T$ dependence for the underdoped
samples with an upturn at low $T$.

The angle-resolved photoemission measurements were carried out 
in two different ARPES systems for all types of samples, 
giving consistent results.  
In one case the photon source is unpolarized with 21.2 eV
photons and a total resolution of 20 meV FWHM, 
while in the other case 22.4 eV linearly-polarized
photons were used with a resolution of 
35 meV FWHM.  The spectra in the figures came from the second
(35 meV) system, while data in Fig.\ \ref{fig3} came from both systems.
The analyzer acceptance angle 
was $\pm 1^\circ$, corresponding to a 
$k$-space window of radius 0.045$\pi/a$ or 
0.037 \AA$^{-1}$.  Base pressures of the
vacuum systems were $4\times 10^{-11}$ torr, and
Fermi energies were determined from a reference Au
film.  Low energy electron diffraction (LEED) 
measurements confirm the quality of the UHV-cleaved 
surfaces and show a Bi-O plane superstructure that
is well known in Bi-based cuprates.

Fermi surface (FS) crossings were determined from at least 
5 $k$-space cuts in 
one octant of each Bi2201 sample.  Two cuts along high
symmetry lines are plotted for three samples with 
different doping levels in Fig.\ \ref{fig1}.  
The data consist of energy distribution curves (EDCs) taken
in the {\it normal state}
at fixed $k$, determined by the angles between the
sample normal and electron analyzer.  A number of 
nearly equivalent criteria can be applied to 
finding the positions of FS crossings 
along $k$-space cuts, such as identifying the point
where the peak intensity decreases by one-half in 
going from occupied to unoccupied states, or where 
the leading edge midpoint of the spectral weight 
comes closest to $E_F$.  The latter (leading edge
midpoint position) is useful because it is also
a measure of a gap in the 
excitation spectrum when it fails to reach 
$E_F$\cite{Shen}.  The measured Fermi surface in our samples is
similar to previous work on overdoped Bi2201\cite{King} showing
a large hole pocket centered on ($\pi$,$\pi$).
We find that over a wide range of doping, there is
little change in the FS and no indication of a FS
topology change.  

The cuts in Fig.\ \ref{fig1} are along
(0,0) to ($\pi$,$\pi$) [$\Gamma Y$ cut] and ($\pi$,0) to 
($\pi$,$\pi$) [$\bar{M}Y$ cut].
The $\Gamma Y$ cuts (left panels) show
the largest energy dispersion and clear FS crossings
in the vicinity of ($0.4\pi,0.4\pi$).  (We plot $\Gamma Y$ instead of 
$\Gamma X$ because there is less complication
from superstructure effects.)  The $\bar{M}Y$ 
cuts (right panels) show much less dispersion, 
but FS crossings 
can still be seen near ($\pi$, 0.25$\pi$).  The
most striking change with doping occurs in the 
lineshape.  As the hole doping increases, the
lineshape becomes much narrower, indicating that
the imaginary part of the excitation's self-energy 
has dramatically decreased.  Only the
overdoped Bi2201 sample approaches Fermi-liquid-like
behavior with well-defined quasiparticle 
excitations.  
The linewidth at ($\pi$,0) is 
especially sensitive to doping, as shown in the
rightmost panel.  The underdoped sample barely has
a peak at all, while the overdoped sample has
a large peak that is nearly resolution-limited. 
Disorder may play a role in the trend of these normal-state spectra,
assuming that Matthiessen's rule holds, 
since the residual resistivity increases from
the overdoped to the underdoped crystals.  The 
residual resistivity ratio (forming R(300 K)/R(0 K)
by extrapolating to 0 K) is 3.7 for the overdoped
case, 2.4 for optimal, and 2.0 for underdoped.
However, the similarity to the trend of linewidth
in Bi2212 suggests that doping is the primary
cause of the linewidth change.  In Bi2212, the 
linewidths of underdoped samples are very broad, but
sharp, resolution-limited peaks still appear below
$T_c$, indicating impurity scattering is not
dominant in these systems\cite{Harris}.

We next turn to the important issue of the
superconducting state gap.  As mentioned above,
the shift in the leading edge midpoint may be used to
characterize an excitation gap.  In the overdoped
case, no gap is observable within our resolution.
This can be seen from the spectra on the FS taken from the
$\Gamma Y$ and $\bar{M}Y$ cuts (Fig.\ \ref{fig2}).
The leading edges of both overdoped spectra coincide.  Since the BCS gap 
for $T_c=8$ K is only 1.5 meV, it is not 
surprising that the gap is too small to be
seen within our error bars of $\pm$2 meV.
In the optimally-doped samples, however, 
we observe a clear and reproducible gap of 10$\pm$2 meV, in rough agreement
with point contact tunneling measurements on ceramic samples of similar
composition\cite{Hudakova}.
The gap is anisotropic, with a maximum on the $\bar{M}Y$ cut [i.e. near
($\pi$,0)] and a minimum consistent with 0 on the $\Gamma Y$ cut 
$45^\circ$ away.  Comparing these two extremes (Fig.\ \ref{fig2})
shows the shift in leading edge position quite clearly.
Furthermore, we have reproduced the same
gap value within error bars on 3 other samples.
Interestingly, the gap persists into the
normal state with no noticeable diminution.
In the underdoped sample, the leading edges
are not as sharp, but a reproducible 
gap of 7$\pm$3 meV maximum magnitude is still evident.

The measured Bi2201 superconducting state lineshape differs from that of
Bi2212 because it lacks the ``peak and dip'' feature.  
In Bi2212, the sharp peak is at $\sim$40 meV binding energy, followed 
by a dip that extends to $\sim$90 meV\cite{Dessau}.
If the features scale in energy with the gap\cite{Varma}, then the 
factor of 3 decrease for Bi2201 would make them difficult to resolve.
It is also possible that impurity scattering plays a role in obscuring
any lineshape change.  Instead of a sharp peak and dip, our spectra show
a moderate linewidth narrowing below $T_c$ (Fig.\ \ref{fig2}).

To explore the gap anisotropy, we took a number of $k$-space 
cuts between the $\Gamma Y$ and $\bar{M}Y$ cuts
of Fig.\ \ref{fig1} on the same samples.  
The leading edge shifts
representing the excitation gap for two optimally doped Bi2201 samples
are plotted vs $0.5|\cos k_xa - \cos k_ya|$ both below and well 
above $T_c$ in Fig.\ \ref{fig3}.
On this plot, a $d_{x^2-y^2}$ gap would be a straight line.  
The curves show considerable
flattening near the origin compared to a pristine $d$-wave gap, suggestive
of either the effect of interlayer tunneling matrix 
elements\cite{CSAS,Yin} or pair breaking due to impurities\cite{dirtyd}.

A central purpose of the present study was to determine whether the 
superconducting state gap anisotropy persists as the
CuO$_2$ planes are increasingly isolated from each other.
Our observation
of a strongly anisotropic gap in Bi2201 indicates that it does, 
since Bi2201 has a large separation between CuO$_2$ 
planes (12.3 \AA\ compared to 3.3 \AA\ for Bi2212), 
a huge $c$-axis resistivity ($\rho_c = 30\: \Omega$ cm at 
50 K\cite{Yoshizaki}), and non-metallic intervening Bi-O layers. 
A theoretical model that may be relevant 
to our findings is the interlayer tunneling (ILT)
model\cite{CSAS}, since it addresses the effect of interplanar coupling on the
gap magnitude and anisotropy.
In the ILT model, the {\it sign} of the order parameter is 
determined by the in-plane pairing kernel, which acts as a 
symmetry breaking field in the space of order parameter 
symmetries\cite{Yin,SudboStrong}.  However, the anisotropy in {\it gap 
magnitude} is dominated by the effect of interlayer matrix 
elements\cite{CSAS,Yin} when $T_c$ is high.  
Thus small $s$-wave or $d$-wave in-plane 
pairing kernels in the presense of strong interlayer coupling would give 
nearly identical results in ARPES gap magnitude measurements.  Weakening
the interlayer coupling more clearly reveals the intrinsic single-plane gap. 
The highly anisotropic gap in Bi2201 is consistent with an underlying
$d_{x^2-y^2}$ symmetry, perhaps with some residual interlayer pair 
tunneling.

Our results mesh well with the tricrystal experiments of Tsuei {\it et al.} 
carried out on the single-plane cuprate Tl2201.  The half-integer 
flux quanta observed in the superconducting state
indicate a sign change in the order parameter\cite{CCTsuei}.  Thus the
intrinsic interaction in a single plane favors $d$-wave 
pairing, while the overall increase in gap magnitude from Bi2201
to Bi2212 by a factor of 3 may indicate that interlayer coupling 
enhances the gap\cite{CSAS,Yin,SudboStrong}.  
Even in Bi2201 the gap magnitude greatly exceeds the BCS prediction 
of 4.4 meV for $T_c=29$K (the leading edge midpoint shift 
tends to underestimate the gap).

It is important
to consider whether disorder and impurities cause the gap to be smaller in
Bi2201 than in Bi2212.  The residual resistivity ratio is 2.4 for 
optimally doped Bi2201, a rather low value that indicates substantial
impurity scattering.  However, some thin films of Dy-doped Bi2212 have
even smaller RRRs, and they show no reduction in gap
magnitude\cite{Harris}.  Thus the lower gap magnitude in Bi2201
is most likely intrinsic.

The measurements of a normal state gap in Bi2201 (Figs.\ \ref{fig2} 
and \ref{fig3}) show that the pseudogap can exist in a one-plane material.  
A pairing enhancement based on
interlayer superexchange J$_\perp$, suggested for bilayer 
materials\cite{MillisMonien,AltUbbens}, will be absent in Bi2201 
because of the large distance between CuO$_2$ planes and 
the geometric frustration induced by the staggering of Cu sites along 
the $c$-axis.  As in the case
of Bi2212\cite{Bi2212ARPES}, the similarity in gap anisotropy and magnitude
above and below $T_c$ suggests that the two gaps are related and that pairing
occurs well above $T_c$.  Further evidence for this point of view
has come from low temperature gap measurements in Bi2212.  
The gap fails to decrease with $T_c$ as
$T_c$ is decreased by underdoping\cite{Harris}.  The persistence of the 
gap in the underdoped Bi2201 sample shows the same anomalous pattern; 
the gap in underdoped samples represents a different energy scale 
from $kT_c$.  Theoretical interpretations of separate energy 
scales for the gap and $kT_c$ have included the idea 
that pairs form at relatively high temperatures but do not become phase
coherent until $T_c$\cite{phasefluc}.  Microscopic theories based 
on spin-charge separation have separate pairings for spin 
and charge excitations\cite{spincharge}, with spinon pairing occurring
in general at a higher temperature than holon pairing.  

The gap measurements on Bi2201 and Bi2212 as a function of doping show
a striking contrast between the effect of lowering $T_c$ by
underdoping and lowering $T_c$ by weakening the interplanar coupling.
The gap (and thus the pairing strength) is relatively 
insensitive to underdoping but drops roughly 
proportionally to the maximum $T_c$ in going from a 
bilayer to single layer material.

While no detailed theory exists for the lineshape evolution with doping,
the change from a narrow peak to a broad continuum
with decreasing hole doping suggests non-Fermi-liquid behavior and
a breakdown of the quasiparticle picture.  
Recently, it was proposed that the peak width is due to the 
strong coupling of electrons to collective excitations with {\bf q} 
peaked at ($\pi,\pi$)\cite{ShenSchrieffer}.  The width
has also been attributed to the decay of the hole into a spinon
and holon\cite{Laughlin}, with the hole lifetime decreasing on 
the underdoped side.  
The rightmost panel of Fig.\ \ref{fig1} shows the lineshape
trend clearly and is insensitive to finite $k$ resolution because the 
dispersion is small near ($\pi$,0).  

In summary, ARPES measurements of the gap and electronic structure of
the one-plane compound Bi2201 open a new window on the occurrence of
superconductivity in a relatively low-$T_c$ member of the cuprate family.
These observations constrain theories by showing that in a system
with very weakly-coupled planes, there is a strongly anisotropic 
superconducting gap and also a pseudogap above $T_c$.


We acknowledge helpful discussions with S. Chakravarty,
S. Strong, S. Kivelson, P. W. Anderson, P. A. Lee, and R. B. Laughlin.
SSRL is operated by the DOE Office of
Basic Energy Sciences, Division 
of Chemical Sciences.  The work at Stanford was supported by ONR Grant
No.\ N00014-95-1-0760; the 
Division of Materials 
Science, DOE; and NSF Grant No.\ DMR 9311566.
H. I. and R. Y. acknowledge a Grant-in-Aid 
from the Ministry of Education, Science, Sports and Culture, Japan.

\pagebreak
\begin{figure}
\vspace*{0.1in}
\caption{Normal state ARPES spectra for underdoped ($T_c<4$ K), 
optimally-doped ($T_c=29$ K), and overdoped 
($T_c=8$ K) Bi2201, measured at $T=100$ K, 60 K, and 60 K, respectively.  
The left panels for each doping show spectra along the $\Gamma Y$ $k$-space 
cut while the right panels show the $\bar{M}Y$ cut.  The rightmost 
panel shows the strong change in linewidth with doping at ($\pi$,0).
}
\label{fig1}
\end{figure}

\begin{figure}
\vspace*{0.1in}
\caption{ARPES spectra at FS crossings for the maximum [near ($\pi, 0.25\pi$)]
and minimum [near ($0.4\pi, 0.4\pi$)] gap for the crystals of Fig.\ \ref{fig1}.
Shifts in the leading edge midpoints indicate an anisotropic gap, as in the
10$\pm$2 meV shift between the arrows for the $T_c=29$ K samples at $T=9$ K.
The spectra are normalized to give the leading edges equal height in
order to show the shifts between them.
}
\label{fig2}
\end{figure}

\begin{figure}
\vspace*{0.1in}
\caption{Leading edge midpoint shifts from $E_F$ for samples A and B, 
indicative of an anisotropic energy gap in the superconducting 
and normal states of optimally-doped Bi2201.
A $d_{x^2-y^2}$ gap is a straight line intercepting the origin of the plot.
}
\label{fig3}
\end{figure}

\begin{references}
\bibitem{Shen}Z.-X. Shen {\it et al.}, Phys. Rev. Lett. {\bf 70}, 1553 (1993).

\bibitem{Campuzano}H. Ding, {\it et al.}, Phys. Rev. B 
{\bf 54}, R9678 (1996).

\bibitem{Bi2212ARPES}D. S. Marshall {\it et al.}, Phys. Rev. Lett.
{\bf 76}, 4841 (1996); A. G. Loeser {\it et al.}, Science {\bf 273}, 
325 (1996); H. Ding {\it et al.}, Nature {\bf 382}, 51 (1996).

\bibitem{Harris}J. M. Harris {\it et al.}, Phys. Rev. B 
{\bf 54}, R15 665 (1996).

\bibitem{pseudogap}For example, see N. P. Ong, Science
{\bf 273}, 321 (1996); B. Batlogg and V. J. Emery, Nature {\bf 382}, 20 (1996); 
B. G. Levi, Physics Today {\bf 49}(6), 19 (1996); and references therein.

\bibitem{MillisMonien} A. J. Millis and H. Monien, Phys. Rev. Lett. 
{\bf 70}, 2810 (1993).

\bibitem{Nameki}H. Nameki {\it et al.}, Physica C {\bf 234}, 255 (1994).

\bibitem{Yoshizaki}R. Yoshizaki, H. Ikeda, L.-X. Chen, M.
Akimatsu, Physica C {\bf 224}, 121 (1994).

\bibitem{King}D. M. King {\it et al.}, Phys. Rev. Lett. 
{\bf 73}, 3298 (1994).

\bibitem{Hudakova}N. Hud\'{a}kov\'{a} {\it et al.}, 
Physica B {\bf 218}, 217 (1996).

\bibitem{Dessau}D. S. Dessau et al., Phys. Rev. Lett. {\bf 66}, 2160 (1991).

\bibitem{Varma}C. M. Varma and P. B. Littlewood, Phys. Rev. B 
{\bf 46}, 405 (1992).

\bibitem{CSAS}S. Chakravarty, A. Sudb\o, P. W. Anderson
and S. Strong, Science {\bf 261}, 337 (1993).

\bibitem{Yin}L. Yin, S. Chakravarty, and P. W. Anderson, preprint.

\bibitem{dirtyd}L. S. Borkowski and P. J. Hirschfeld, Phys. Rev. B {\bf 49}, 
15404 (1994).

\bibitem{SudboStrong}A. Sudb\o\ and S. P. Strong, Phys. Rev. B {\bf 51}, 
1338 (1995).

\bibitem{CCTsuei}C. C. Tsuei {\it et al.}, Science {\bf 271}, 329 (1996). 

\bibitem{AltUbbens}M. U. Ubbens and P. A. Lee, Phys. Rev. B 
{\bf 50}, 438 (1994); 
B. L. Altshuler, L. B. Ioffe, and A. J. Millis, Phys. Rev. B 
{\bf 53}, 415 (1996).

\bibitem{phasefluc}V. J. Emery and S. A. Kivelson, Nature {\bf 374}, 434 (1995);
S. Doniach and M. Inui, Phys. Rev. B {\bf 41}, 6668 (1990).

\bibitem{spincharge}P. W. Anderson and S. P. Strong, Chinese J. Phys. {\bf 34},
159 (1996); P. W. Anderson, J. Phys. Cond. Matt. {\bf 8}, 10083 (1996);
X.-G. Wen and P. A. Lee, Phys. Rev. Lett. {\bf 76}, 503 (1996);
V. J. Emery, S. A. Kivelson, and O. Zachar (preprint, cond-mat/9610094).

\bibitem{ShenSchrieffer}Z.-X. Shen and J. R. Schrieffer, 
Phys. Rev. Lett. {\bf 78}, 1771 (1997).

\bibitem{Laughlin}R. B. Laughlin (preprint, supr-con/9608005).


\end{references}
\end{document}